\documentclass[%
 aip,
 sd,%
 amsmath,amssymb,
 reprint,%
]{revtex4-1}

\usepackage{graphicx}% Include figure files
\usepackage{dcolumn}% Align table columns on decimal point
\usepackage{bm}% bold math
\begin{document}

\preprint{AIP/123-QED}

\title[Hindered nematic alignment of hematite spindles in viscoelastic matrices]{Hindered nematic alignment of hematite spindles in viscoelastic matrices}% Force line breaks with \\

\author{A. Nack}
\affiliation{ 
Institut f\"ur Chemie, Abteilung Physikalische Chemie, Universit\"at Rostock, Albert-Einstein-Str. 3\,a, D-18055 Rostock%\\This line break forced with \textbackslash\textbackslash
}%
\author{J. Seifert}
\affiliation{%
Department Chemie, Institut f\"ur Physikalische Chemie, Universit\"at zu K\"oln, Luxemburger Str. 116, D-50939 K\"oln
}%
\author{C. Passow}
\affiliation{%
Deutsches Elektronen-Synchrotron (DESY), Notkestrasse 85, D-22607 Hamburg 
}%
\author{J. Wagner}
\affiliation{ 
Institut f\"ur Chemie, Abteilung Physikalische Chemie, Universit\"at Rostock, Albert-Einstein-Str. 3\,a, D-18055 Rostock%\\This line break forced with \textbackslash\textbackslash
}%

\date{\today}% It is always \today, today,

\begin{abstract}
The viscoelastic behavior of composites consisting of spindle-shaped hematite particles in
poly-N-isopropylacrylamide hydrogels is investigated both, by means of rheological oscillatory shear experiments, and
the field-induced alignment of these mesoscale, anisotropic particles in external magnetic fields.
Due to their magnetic moment and magnetic anisotropy hematite spindles align with their long axis perpendicular to
the direction of an external magnetic field.
The field induced torque acting on the magnetic particles leads to an elastic deformation of the hydrogel
matrix. Thus, the field-dependent orientational distribution functions of anisotropic particles acting as microrheological 
probes depend on the elastic modulus of the hydrogel matrix. The orientational distribution functions are determined by means of Small Angle X-ray Scattering
experiments in presence of external magnetic fields. 
With increasing elasticity of the hydrogels, tuned via the polymer volume fraction and the crosslinking density, the field-induced
alignment of these anisotropic, magnetic particles is progressively hindered. The microrheological results are in accordance to 
macrorheological experiments indicating increasing elasticity with increasing flux density of an external field. 
\end{abstract}

%Use showkeys class option if keyword
\maketitle

\section{Introduction} Since the pioneering work of Payne et al. \cite{Payne1971} the influence of incorporated nanoparticles on the mechanical properties of elastomers has attracted considerable scientific interest. The macroscopic viscoelastic properties of such composites are strongly influenced by interactions between embedded particles and the polymer network. In many cases, polymer adsorption at the particle surface leading to additional junctions between polymer chains is discussed as the microstructural reason for enhanced mechanical stability \cite{Sternstein2002}.\\
If embedded particles carry a magnetic moment, due to the interaction between particles and magnetic fields and particle-matrix interactions, the mechanical properties of such ferrogels are influenced by the flux density and direction of external magnetic fields. Field-induced changes of macroscopic mechanical, magnetic and optical  properties and their relation
to changes of structure and dynamics at the mesoscale are currently in the focus of scientific interest, both, from the viewpoint of fundamental and applied sciences. The scientific significance of such smart materials is driven by their high potential for technical and biomedical applications \cite{Ramanujan2006,Ghosh2010,Deng2006}.
For ferrogels, the thermoresponsive hydrogel poly(N-isopropylacrylamide) (pNIPAM) is one of the commonly chosen and well characterized matrix systems for the preparation of such smart materials \cite{Pelton2000}. At its lower critical solution temperature (LCST) of $\sim$ 32 $^\circ\,$C it undergoes a phase transition from a soluble to an insoluble state, accompanied by the collapse of the polymer structure. Due to the volume-phase transition close the human body temperature, medical applications of composites of pNIPAM and magnetic nanoparticles especially in the area of drug targeting techniques arise \citep{Chang2013,Schild1992,Purushotham2009}.\\
In up to now described ferrogels based on pNIPAM or other soft matrices, e.g., gelatine or polyethylen glycole (PEG), spherical magnetic nanoparticles such as maghemite ($\gamma$-\,Fe$_2$O$_3$) or cobalt ferrite (CoFe$_2$O$_4$) are incorporated \cite{Nakao2014,Backes2015}. These spherical magnetic nanoparticles are used as microrheological probes
to investigate the viscoelastic properties of such composites by means of AC-susceptometry \citep{Roeben2014,Remmer2015}. In this contribution, spindle-shaped magnetic hematite nanoparticles with tunable aspect ratio are used as microrheological probes to investigate elastic and viscous properties of ferrogels.\\
If shape anisotropic, magnetic particles are embedded in viscoelastic matrices, their mesoscale structure and dynamics and the related macroscopic properties depend on the particles' orientation as an additional degree of freedom.  Distortions of the matrix either induced by interaction of the particles with an external field or mechanical stress exerted  on the composites influence, both, the spatial and orientational distribution of the particles. In addition, shape anisotropic nanoparticles as mesogens are constituents of lyotropic, mineral liquid crystals (LCs). Both, discotic and calamitic mineral LC phases, consisting, e.g., of boehmite ($\gamma$-AlO(OH)) plates or goethite ($\alpha$-FeO(OH)) as well as silica rods, are described \cite{Lekkerkerker2012,Gabriel2000,Murphy2016}. Already in 1949, Onsager predicted the formation of a nematic phase by mineral colloids with the decrease in excluded volume compensating a loss of orientational entropy \cite{Onsager1949}.  
The dynamic behavior of discotic colloidal particles in the shear flow of a Newtonian liquid \cite{Lettinga2012} as well as
in the nematic phase of 5CB\cite{Leheny2012} is investigated. The phase behavior of ellipsoidal magnetic particles consisting of silica-coated hematite particles in external magnetic fields is investigated by Martchenko et al. \cite{Martchenko2016}. The interaction of diluted aqueous suspensions of such ellipsoidal core-shell particles with an external magnetic field has been described by Reufer et al.\citep{Reufer2010,Reufer2012}.\\
Despite the volume fraction of the hematite spindles in the gels investigated here is unsufficient to induce the formation of a nematic order caused by excluded volume, due to the interaction of the spindles' magnetic moments, in first approximation perpendicular to the spindle axis, a field-induced isotropic-nematic phase transition can be observed: in moderate external fields, hematite spindles align with their long axes perpendicular to the field direction as observed by means of  Small Angle X-Ray Scattering (SAXS) and optical birefringence \cite{Markert2011,Dagallier2010,Dagallier2012}. Opposite to the studies
of Dagallier et al. and Roeder et al. \cite{Roeder2014,Roeder2015}, using pNIPAM grafted at the particle surface, in the present work, hematite spindles are not rigidly attached to, but confined in cavities of a inter-crosslinked polymer hydrogel network formed by pNIPAM.  We investigate the composites' mesostructure by means of SAXS and their complex macroscopic rheological properties by means of oscillatory shear experiments, both in dependence on the magnetic flux density of an external field. 

\section{SAXS of aligned anisotropic particles}

Hematite spindles interact with an external magnetic field due to  both, a permanent magnetic moment nearly perpendicular to their rotation axis\cite{Golosovsky:2007,Lemaire2004_2}, which is parallel to the trigonal crystallographic direction, and an induced moment proportional to their negative, magnetic anisotropy $\Delta \chi=\chi_{\parallel}-\chi_{\perp}$
\begin{equation}
\label{eq:energy}
V(\vartheta_{\rm{P}})= -\mu B \cos{\vartheta_{\mu}} - \frac{\Delta \chi V_{\rm P} B^2}{2 \mu_0} \cos^2 \vartheta_{\rm{P}}\,.
\end{equation}
$\chi_\parallel$ and $\chi_\perp$ denote the magnetic susceptibilities parallel and perpendicular to the spindles' director defined by its axis of revolution. The Zeeman contribution of the permanent moment is proportional to the cosine of the angle $\vartheta_\mu$ enclosed between the direction of the field with the magnetic flux density $B$ and the direction of the magnetic moment with the modulus $\mu$. The contribution of the induced magnetic dipole is proportional to the particle volume $V_{\rm P}$ and the square $B^2$ of the flux density. Its angular dependence can be described by $\cos^2\vartheta_{\rm P}$, where $\vartheta_{\rm P}$ denotes the angle between field direction and particle director. The direction of the permanent magnetic moment is related to the particle director by an offset $\vartheta_{\rm off}$ and can be obtained by an Eulerian tranformation in dependence on the Eulerian angles $\vartheta_{\rm P}$ and $\chi_{\rm P}$.

Using \eqref{eq:energy}, a Boltzmann ansatz for the orientational distribution function (ODF)
of the particles can be written as 
\begin{align}
\label{eq:Boltzmann_ODF}
p(\vartheta_{\rm{P}},\chi_{\rm P})&= \frac{1}{Z}\exp \left[\dfrac{\mu B}{k_{\rm{B}}T}\cos\left[\vartheta_\mu(\vartheta_{\rm P},\chi_{\rm P})\right]\right] \notag \\
&\cdot\exp \left[\dfrac{\Delta\chi V_{\rm P} B^2}{2\mu_0 k_{\rm B}T}\cos^2\vartheta_{\rm P}\right]
\end{align}
with the thermal energy $k_{\rm B}T$ and the partition function $Z$ as the integral of the exponential over the Eulerian angles $\varphi_{\rm P},\vartheta_{\rm P}$, and $\chi_{\rm P}$. Here, particle interactions between the hematite spindles are neglected due to their large interparticle distances at very small volume fractions in the order of $10^{-3}$ compared to the interactions between individual particles with the external field.\\
Employing the ODF as normalized probability density and the form factor $P(Q,\vartheta_{\rm Q})$\cite{Markert2011}, where $Q$ and $\vartheta_{\rm Q}$ are the modulus and the angle of the particle director with respect to the direction of the scattering vector, the scattered intensity can be expressed as 
\begin{align}
I&(Q,\vartheta_{\rm Q})=\int\limits_0^{2\pi}\int\limits_0^{1}\int\limits_0^{2\pi} p(\vartheta_{\rm{P}},\chi_{\rm{P}})\\ \nonumber 
&\qquad \times P\left[Q,\gamma(\vartheta_{\rm Q},\varphi_{\rm Q};\vartheta_{\rm{P}},\varphi_{\rm{P}})\right]
\rm{d}\varphi_{\rm{P}}\,\rm{d}\cos{\vartheta_{\rm{P}}\,\rm{d}\chi_{\rm{P}}},
\end{align}
where the $\gamma(\vartheta_{\rm Q},\varphi_{\rm Q}; \vartheta_{\rm P},\varphi_{\rm P})$ denotes the cosine of the angle enclosed between the direction $(\vartheta_{\rm Q},\varphi_{\rm Q})$ of the scattering vector and the particle director
$(\vartheta_{\rm P},\varphi_{\rm P})$.\\
From the simultaneous analysis of scattering data of a hematite suspension in water as a Newtonian liquid, containing hematite spindles with an aspect ratio $\nu=4.0$ and an average length of the long axis of 200 nm, in dependence on the modulus of the scattering vector and its direction relative to the external field as well as in dependence on the flux 
density, the permanent dipole moment is determined as $\mu=6.225_2\times 10^{-19}\, {\rm A}\,{\rm m^2}$. For the magnetic anisotropy $\Delta\chi=-3.054_1\times 10^{-3}$ and for the offset $\vartheta_{\rm off}=78.6_1^\circ$ is obtained.\\
Due to the proportionality to $B^2$, the induced dipole will dominate at sufficiently high flux densities. For $B>0.2\,{\rm T}$ the contribution of the induced dipole exceeds that of the permanent dipole for the particles used in these experiments.\\
The nematic order parameter $S_2$ calculated from the field-dependent ODF of the hematite spindles by canonical averaging of the second Legendre polynomial \cite{deGennes1993}
\begin{equation}
\label{eq:s2_nematic}
S_2=\frac{1}{2}\langle 3\cos^2\vartheta_{\rm{P}} -1\rangle
\end{equation}
is used to quantify the alignment of the magnetic nanoparticles as a function of the flux density. If $\vartheta_{\rm P}$ denotes the angle between external field and the long particle axis, for complete alignment of the particles perpendicular to the external field, $S_2=-1/2$ is expected, while this order paramter vanishes for a random alignment.\\

\section{Experimental}

\begin{figure}
\centering
  \includegraphics[width=0.4\linewidth]{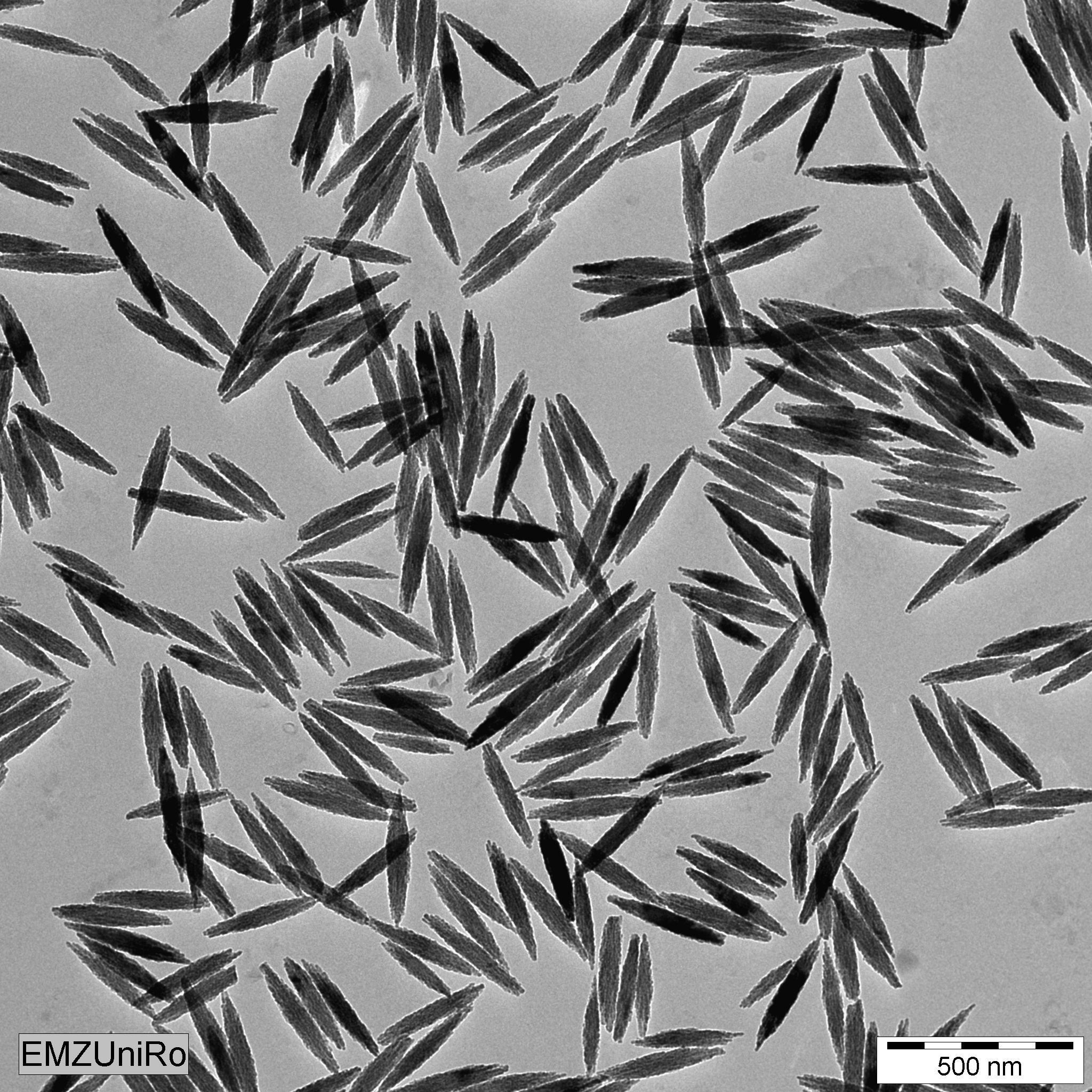}
  \hspace*{0.2cm}
  \includegraphics[width=0.4\linewidth]{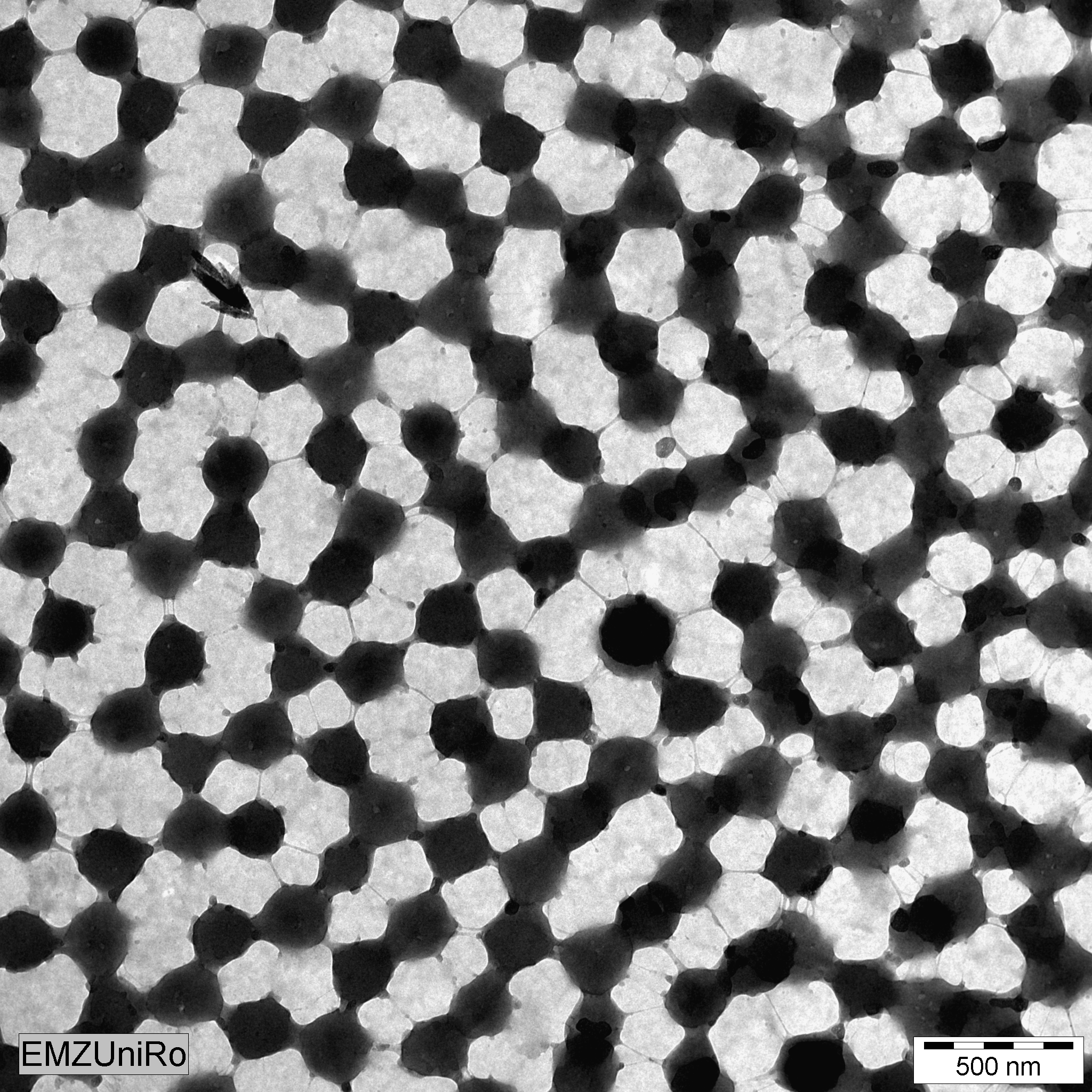}
  \caption{Transmission electron micrograph of hematite particles (lhs) and GA crosslinked pNIPAM hydrogel (rhs)}
  \label{fig:TEM}
\end{figure}

Spindle-shaped hematite ($\alpha$-Fe$_2$O$_3$) particles (FIG.\,\ref{fig:TEM}, lhs) are prepared by the thermolysis 
\begin{equation}
\mbox{FeCl}_3 + 3 \mbox{H}_2\mbox{O} \xrightarrow[\mbox{NaH}_2\mbox{PO}_4]{\Delta,\,\,48\,{\rm h}\\
} \mbox{Fe}_2\mbox{O}_3 + 6 \mbox{HCl}
%reaktionsbedingungen: \longrightarrow[48 h]{\mbox{NaH}_2\mbox{PO}_4} 
\end{equation}
of an aqueous ferric chloride ($\mbox{FeCl}_3$) solution in presence of sodium dihydrogen phosphate ($\mbox{NaH}_2\mbox{PO}_4$) as desribed by Ozaki et al. \cite{Ozaki1984}.
An aqueous solution of 0.25 mmol/L $\mbox{NaH}_2\mbox{PO}_4$ \textit{(Sigma-Aldrich)} is heated to the boiling point and reflux. Subsequently 0.02 mol/L $\mbox{FeCl}_3$ \textit{(J.T. Baker)} is added and the reaction is kept under reflux for 48 hours. In the processing step, the suspension is centrifuged stepwise and the remaining centrifugate washed with deionized water.\\ 
Since the distance $d_{\rm{O-O}}$=2.5\,\AA\,\ of oxygen atoms in phosphate anions fits better to the distance of iron atoms $d_{\rm{Fe-Fe}}$=2.29\,\AA\,\, at surfaces parallel than perpendicular  to the trigonal axis with  $d_{\rm{Fe-Fe}}$=2.91\,\AA , phosphate regioselectively adsorbs at surfaces parallel to the trigonal axis \cite{Sugimoto1998}. 
Due to this regioselective adsorption and the preferred growth in the direction of the trigonal axis,  the concentration of phosphate anions tunes the aspect ratio of hematite particles.\\
The poly(N-isopropylacrylamide) hydrogel matrix \cite{Hu2003} is prepared by means of free radical polymerization of N-isopropylacrylamide (NIPAM) in water in the presence of glutaric aldehyde (GA) as a crosslinker  and potassium persulfate ($\mbox{K}_2\mbox{S}_2\mbox{O}_8$) as free-radical initiator. 
Water heated to 80\,$^{\circ}$\,C is for one hour continuously stirred in a nitrogen atmosphere to remove oxigen.  An aqueous solution of 50 wt\% GA \textit{(Fluka)}, NIPAM \textit{(TCI)} and SDS \textit{(Merck)} are added to the mixture as given in TABLE\,\ref{tab:pnipam}. After all additives completely have been dissolved in water, an aqueous solution of $\mbox{K}_2\mbox{S}_2\mbox{O}_8$ \textit{(Sigma-Aldrich)} is added and the reaction mixture is kept under continous stirring  at 80\,$^{\circ}$\,C in a nitrogen atmosphere for 4 hours. Subsequently, the resulting gel is dialyzed against destilled water for at least one week and reduced to a volume of 100 ml via vacuum distillation.\\

\begin{table}
\caption{Overview of amount of chemicals used to prepare pNIPAM hydrogels of two different crosslinking densities $\chi$. }\begin{tabular}{p{1.6cm}p{3cm}p{3cm}}
\hline
\hline
 & pNIPAM $\chi=0.1$& pNIPAM $\chi=0.05$ \\
\hline
NIPAM & 0.18 mol/L & 0.18 mol/L \\ 
GA & 0.58 mmol/L & 0.29 mmol/L \\ 
SDS & 0.28 mmol/L & 0.28 mmol/L \\ 
$\mbox{K}_2\mbox{S}_2\mbox{O}_8$ & 1.48 mmol/L & 1.48 mmol/L \\ 
H$_2$O & 0.25 L & 0.25 L \\
\hline
\hline
\end{tabular}
\label{tab:pnipam}
\end{table}

Due to GA, a bifunctional $\alpha-\omega$-dialdehyde, a hydrogel consisting of polymer spheres linked by polymer chains, is obtained (FIG.\,\ref{fig:TEM}, rhs). The viscosity of the hydrogel matrix can be  adjusted by the polymer volume fraction $\varphi=V(\rm{monomer})/V(\rm{total})$. For monomer concentrations of $c_{\rm{NIPAM}}=0.18 $ mol/L and a polymer density $\rho_{\rm{NIPAM}}\approx 1\,\rm{g}/\rm{cm}^3$ this results in a polymer volume fraction $\varphi=0.05$. By diluting the sample with water, lower polymer volume fractions are achieved. All polymer volume fractions are chosen above the critical volume fraction of pNIPAM preventing a phase separation to water and a gel phase to ensure macroscopically homogeneous particle-gel composites.The crosslinking ratio is determined by the molar ratio  
\begin{equation}
\chi=\dfrac{n_{\rm GA}}{n_{\rm GA}+n_{\rm NIPAM}}.
\end{equation}
%n_{\mbox{cross}} = \frac{n(\mbox{glutaric aldehyde})}{n(\mbox{N-isopropylacrylamide})+n(\mbox{glutaric aldehyde})}.
In the following, samples with  crosslinking ratios up to $\chi$= 0.1 are investigated.

Hematite-hydrogel composites are prepared by mixing hydrogels and aqueous hematite suspensions and stirring until homogeneity is achieved as verified by optical transmission. The homogeneity in addition is verified by SAXS experiments illuminating different sample volumes of the sample resulting in reproducible scattering data.\\
The X-ray scattering experiments are performed at the high brilliance beamline ID02 at the European Synchrotron Facility (ESRF) in Grenoble. The incident energy is chosen to 12 keV ($\lambda$=0.997 \AA) with a detector distance of $d$ = 20\,m. A magnetic field with flux densities between 0.03 T and 1.5 T perpendicular to the primary beam is applied by rare earth magnets with a variable pole distance. The illuminated sample volume is with $1\times0.1\times0.1$ mm$^3$ significantly smaller than the pole size ($100\times50$ mm$^2$) of the magnets. Hence,  the magnetic field in the illuminated sample volume is gradient free and highly homogeneous.\\
Rheological investigations are executed by using a MCR 302 rheometer by Anton-Paar in plate-plate geometry with a diameter of $d=16\,{\rm mm}$ applying the MRD (magneto-rheological measuring cell) setup. Hence within the investigated volume flux densities of up to 1.0\, T can be applied during measurements. The temperature is monitored via a PT 100 sensor in the center of the bottom plate  in immediate vicinity of the sample.
During the field-dependend experiments a temperature stability better than $\Delta T=\pm 0.15\,{\rm K}$ is achieved. The direction of the external magnetic field is parallel to the direction of the applied shear gradient.\\
The viscous, but still flowing samples are applied by means of a pipette with enlarged tip in the center of the lower measuring plate. Before starting the oscillatory shear experiment, the sample is exposed to a rotational shear stress for one minute at a shear rate of $\dot\gamma=10\,\rm{s}^{-1}$, to ensure homogeneity of the ferrogel in the gap between both measuring plates. Experiments are performed at constant temperature of 20$^{\circ}$ C.\\ 
%in accordance to the on-site hutch temperature of the SAXS experiment at ID02, since the experimental magnetic setup available does not include temperature dependent experiments.
Storage- and loss moduls $G'$ and $G''$ of pNIPAM hydrogels and therewith the orientation of the hematite spindles within the matrix can be influenced by the temperature as an external parameter due the characteristic pNIPMAM coil-to-globuli transition. However, temperature-dependent rheological measurements concerning the mechanics of the matrix are not presented in this work, since the Small Angle X-ray Scattering experiments are performed at the identical temperature of  20\,$^{\circ}$\,C, too.

\section{Results and Discussion}

The hematite particles are topologically characterized by means of TEM and SAXS. SAXS experiments, in addition to the topological characterization give access to the orientational distribution function of the particles in dependence on the flux density of an external magnetic field. To investigate correlations between the mesoscale structure and macroscopic properties of the composites, oscillatory shear experiments using the same samples and experimental conditions used for the scattering experiments are performed. While static Small Angle X-ray Scattering experiments give insights into the structure at mesoscopic length scales, rheological experiments characterize macroscopic, viscoelastic properties of the samples.

\subsection{Topology of the spindle-shaped hematite particles}

\begin{figure}
\centering
\includegraphics[scale=0.4]{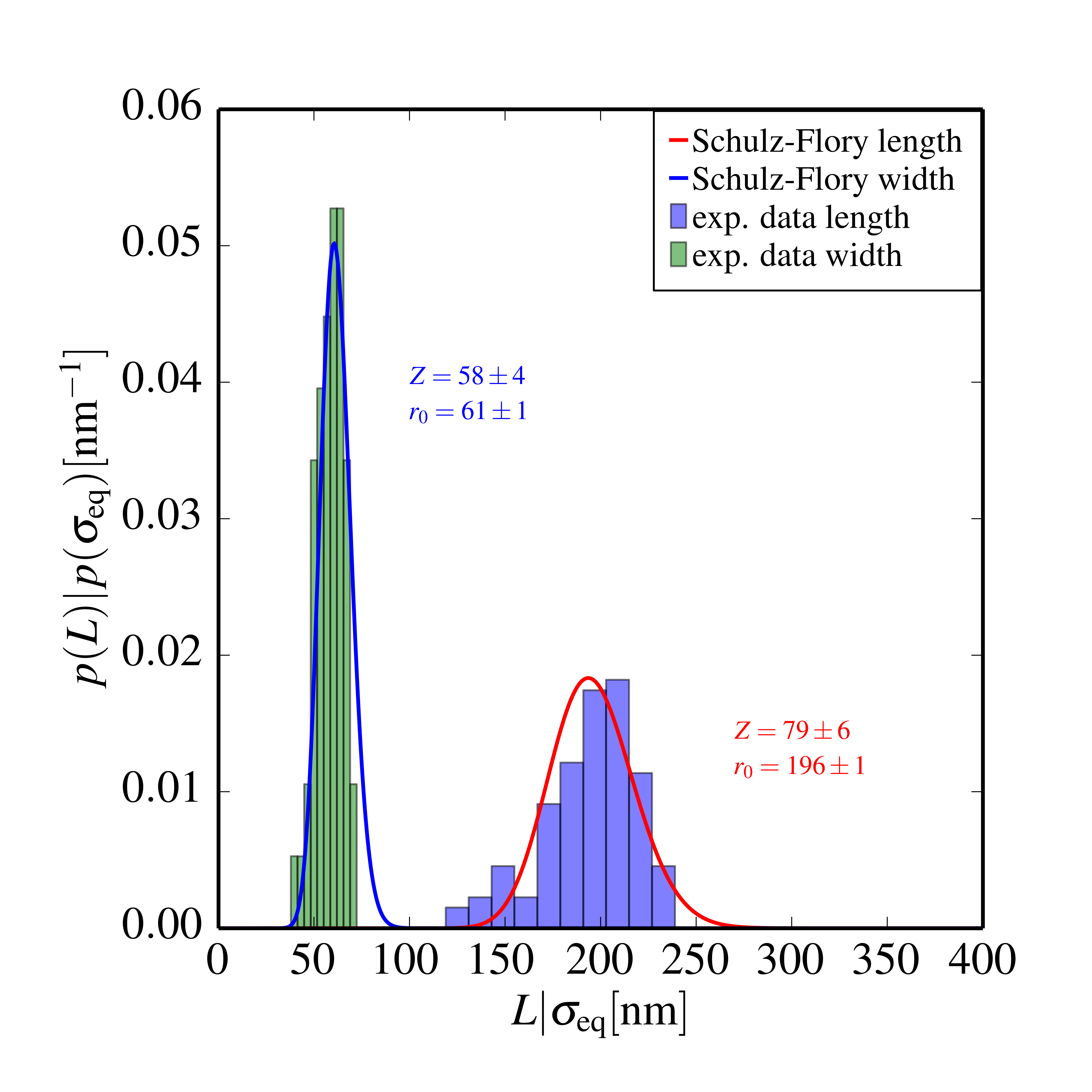}
\caption{Histogram of size distributions of the particles length (blue) and width (green) as well as Schulz-Flory fit (solid lines) to the distributions. $Z$ is related to the systems' polydispersity and $r_0$ denotes the mean of the distribution.}
\label{fig:size_dist}
\end{figure}

In this work, spindle shaped hematite nanoparticles serve as probe for the investigation of a viscoelastic pNIPAM hydrogel network. 
The distribution of particle lengths and diameters is determined from electron micrographs (FIG.\,\ref{fig:TEM}, lhs) of several ensembles of hematite spindles. The resulting, normalized histograms  (FIG.\,\ref{fig:size_dist}) are described via a Schulz-Flory distribution\cite{Markert2011}, the optimum parameters of which are compiled in TABLE\, \ref{tab:size_dist}.\\
These topological parameters, in addition, are obtained by a simultaneous least-squares fit of  the experimental SAXS data at 27 logarithmically equidistant flux densities in the range of $3\times 10^{-3} \,\rm{T} \leq B \leq 1.5 \,\rm{T} $, depending on the modulus and the direction of the scattering vector relative to the external field and the magnetic flux density (compare FIG.\,\ref{fig:false_color} and \ref{fig:intensity}) using more than $2\times 10^5$ symmetrically independent data points. For numerical reasons, a model with a particle form factor of spindles with a constant aspect ratio,  underlying a Schulz-Flory distribution of equatorial diameters is applied.The optimimum parameters for equatorial diameter $\sigma_{\rm eq}$, aspect ratio $\nu$ and $Z$-parameter of the Schulz-Flory distribution are compiled in TABLE\,\ref{tab:size_dist}, as well. The polydispersity defined as the square root of the reduced variance is related to the $Z-$parameter of the Schulz-Flory distributions by 
\begin{equation}
P=\left(\dfrac{\left\langle \sigma_{\rm eq}^2\right\rangle-\left\langle \sigma_{\rm eq}\right\rangle^2}{\left\langle \sigma_{\rm eq}\right\rangle^2}\right)^{1/2}=\left(\frac{1}{Z+1}\right)^{1/2}.
\end{equation}
The mean aspect ratio determined from TEM in TABLE\,\ref{tab:size_dist} is the average of the individual aspect ratios from single particles which is larger than the ratio $\langle L\rangle/\langle \sigma_{\rm eq}\rangle$. As already observed in previous work \cite{Markert2011}, the aspect ratio determined from SAXS exceeds the one determined by means of TEM. Opposite to analysis of TEM using at maximum some hundred particles, SAXS gives access to more representative  ensemble exceeding the
ensemble practically accessible by TEM by several orders of magnitude. In addition, by TEM only the projection of the long particle axis to the image plane is visible, which is systematically shorter than the actual size, if the particles are not exactly perpendicular to the electron beam axis. Furthermore, in case of overlapping particles the exact determination of the particle boundary is difficult, especially for magnetic particles usually producing electron-optical artifacts in the boundary areas of the particles.

\begin{table}
\caption{Summary of fit parameters $\sigma_{\rm{eq}}$, $L$, $Z$ and $\nu$ from Schulz-Flory fits to TEM evaluated size distribution and SAXS data as well as calculated polydispersity $P$ determined for the hematite sample used in this work.}
\begin{tabular}{cccccccc}
\hline 
\hline 
 & \textbf{$\sigma_{\rm{eq}}$} & \textbf{$L$ } & \textbf{$\nu$ } & \textbf{$Z(\sigma_{\rm{eq}})$}& \textbf{$Z(L)$} &\textbf{$P(\sigma_{\rm{eq}})$}&   \textbf{$P(L)$}  \\ 
\hline 
TEM & 61 & 196 & 3.4 & 58 & 79 & 0.13 & 0.11 \\ 
%\hline 
SAXS & 54 & 212 & 4 & 46 & 46 & 0.14 & 0.14 \\ 
\hline 
\hline 
\end{tabular} 
\label{tab:size_dist}
\end{table}

\subsection{Small Angle X-ray Scattering}

\begin{figure}
\centerline{
  \includegraphics[width=0.2\linewidth]{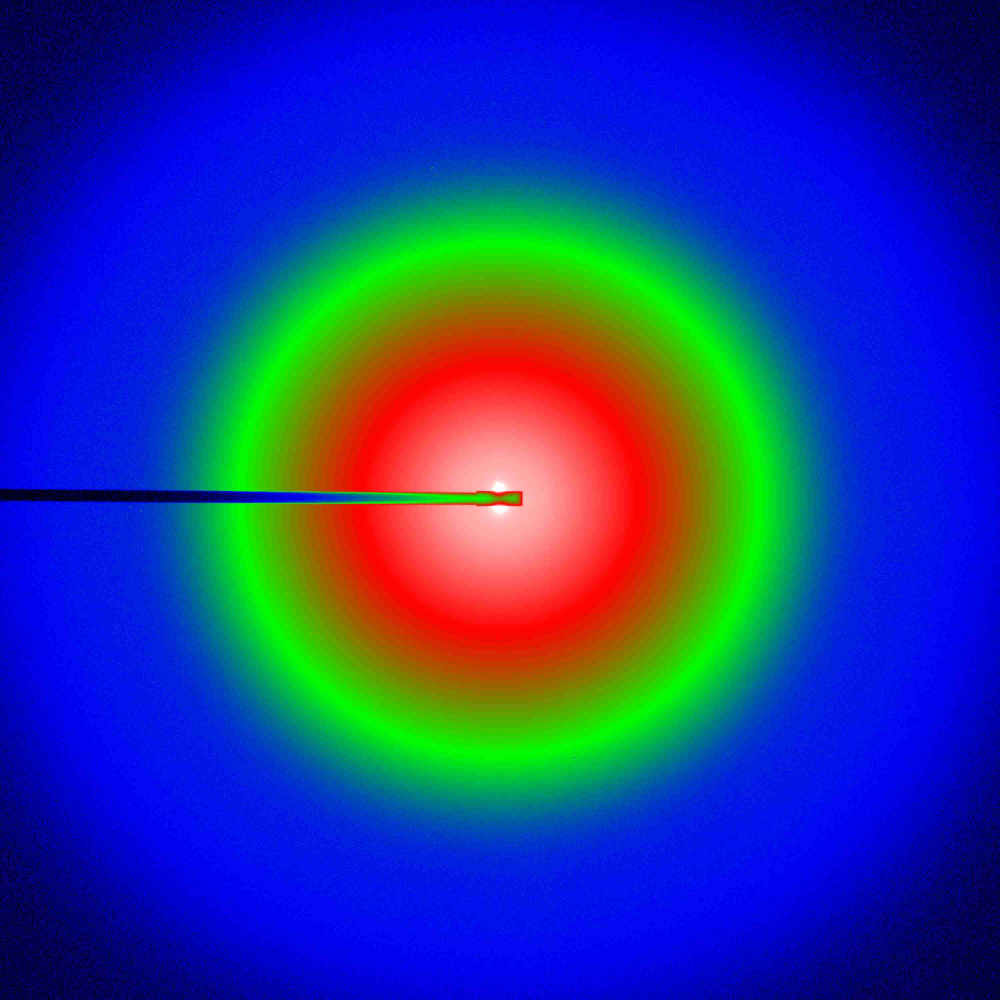}\hfill
  \includegraphics[width=0.2\linewidth]{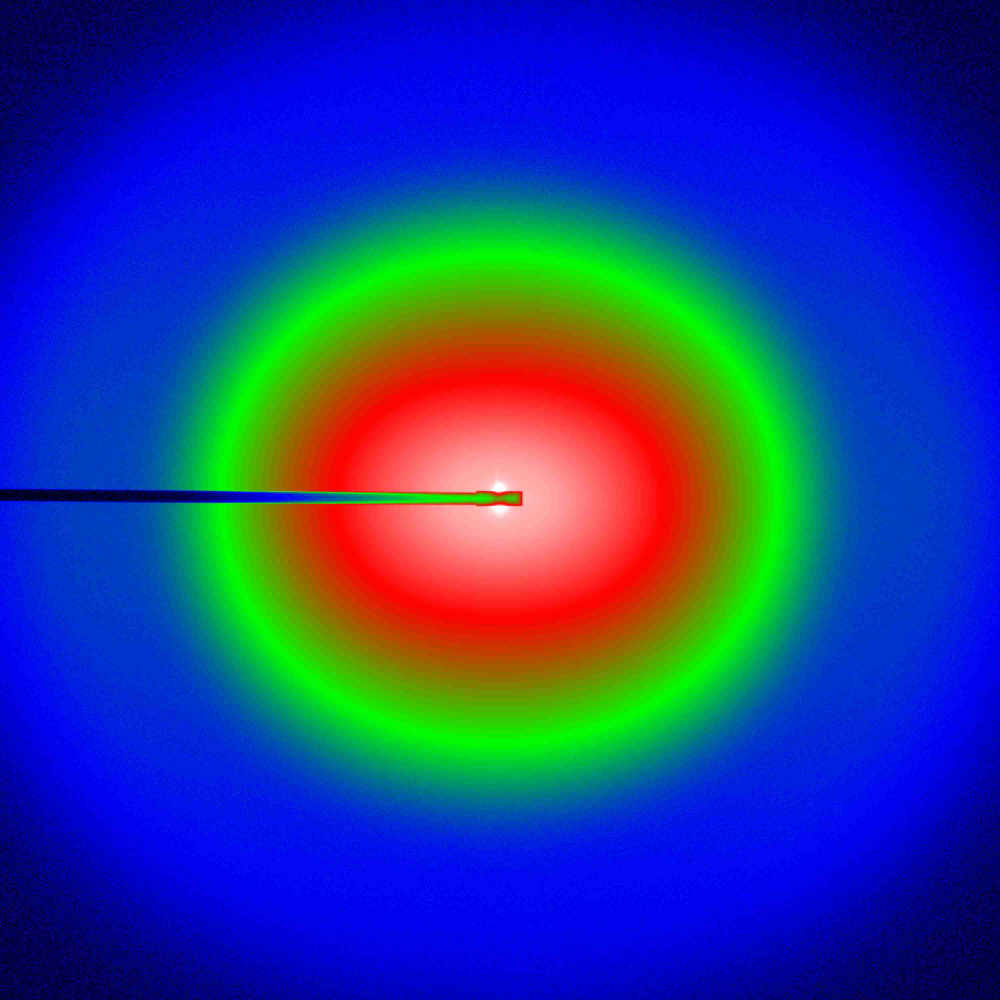}\hfill
  \includegraphics[width=0.2\linewidth]{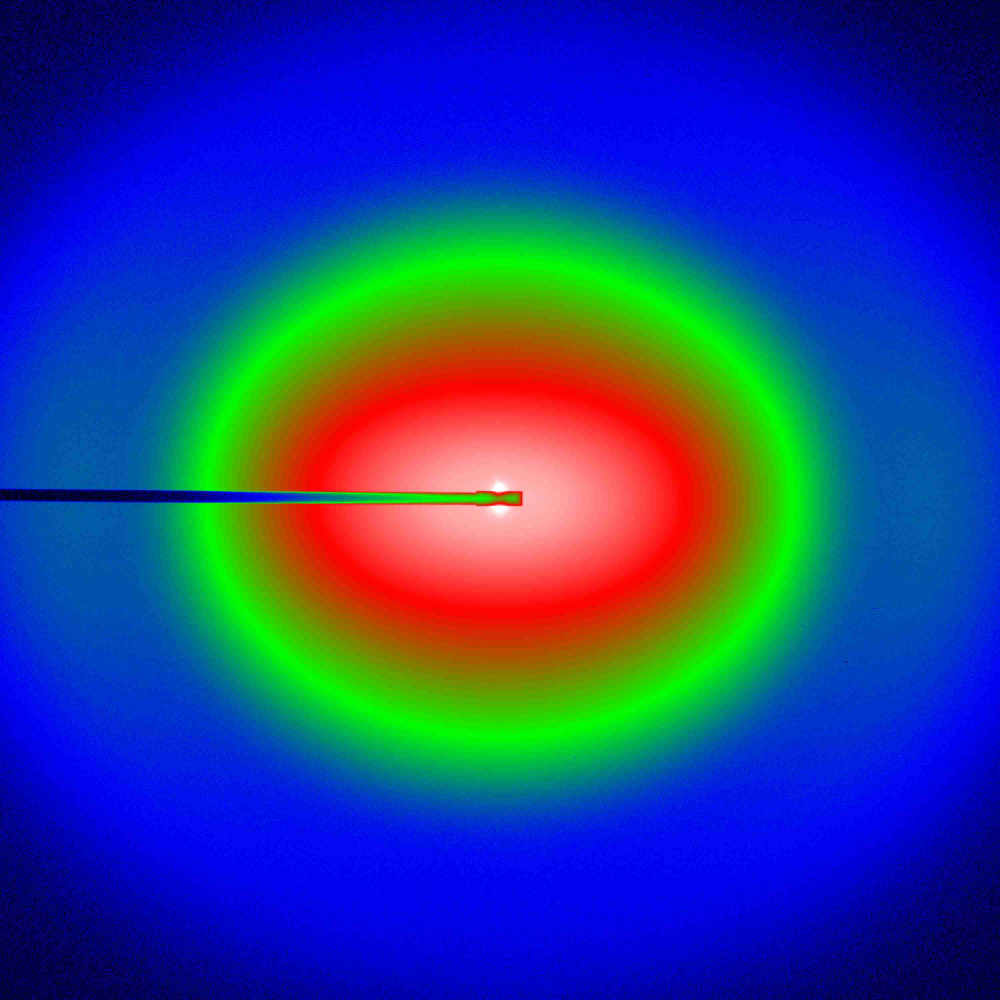}\hfill
  \includegraphics[width=0.2\linewidth]{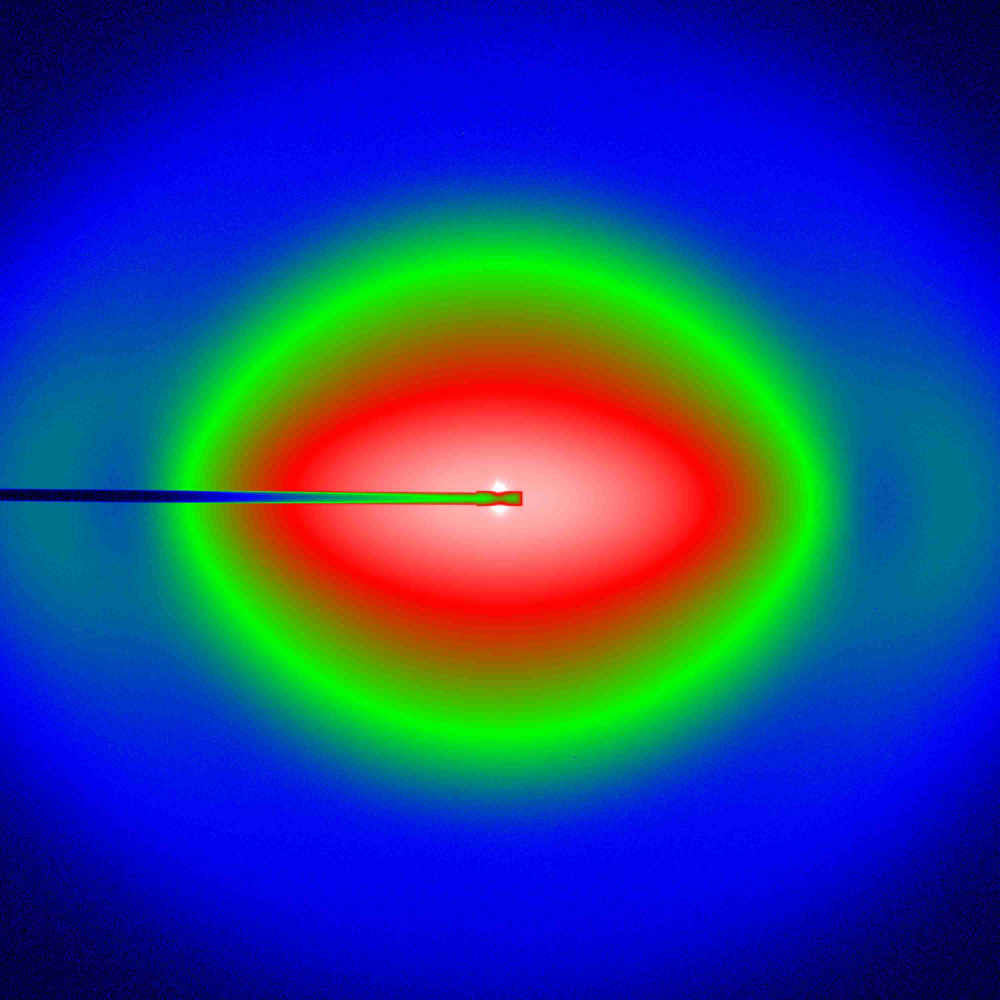}\hfill
  \includegraphics[width=0.2\linewidth]{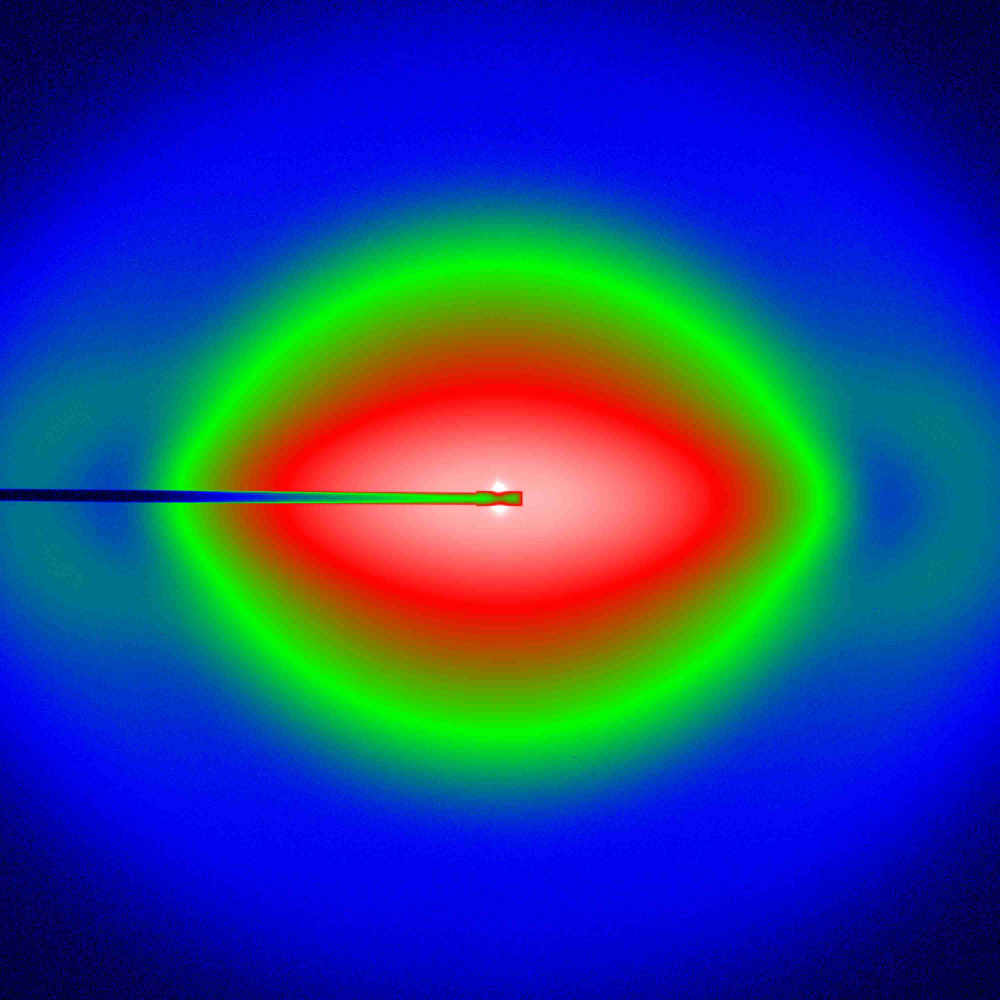}}
\centerline{
  \includegraphics[width=0.2\linewidth]{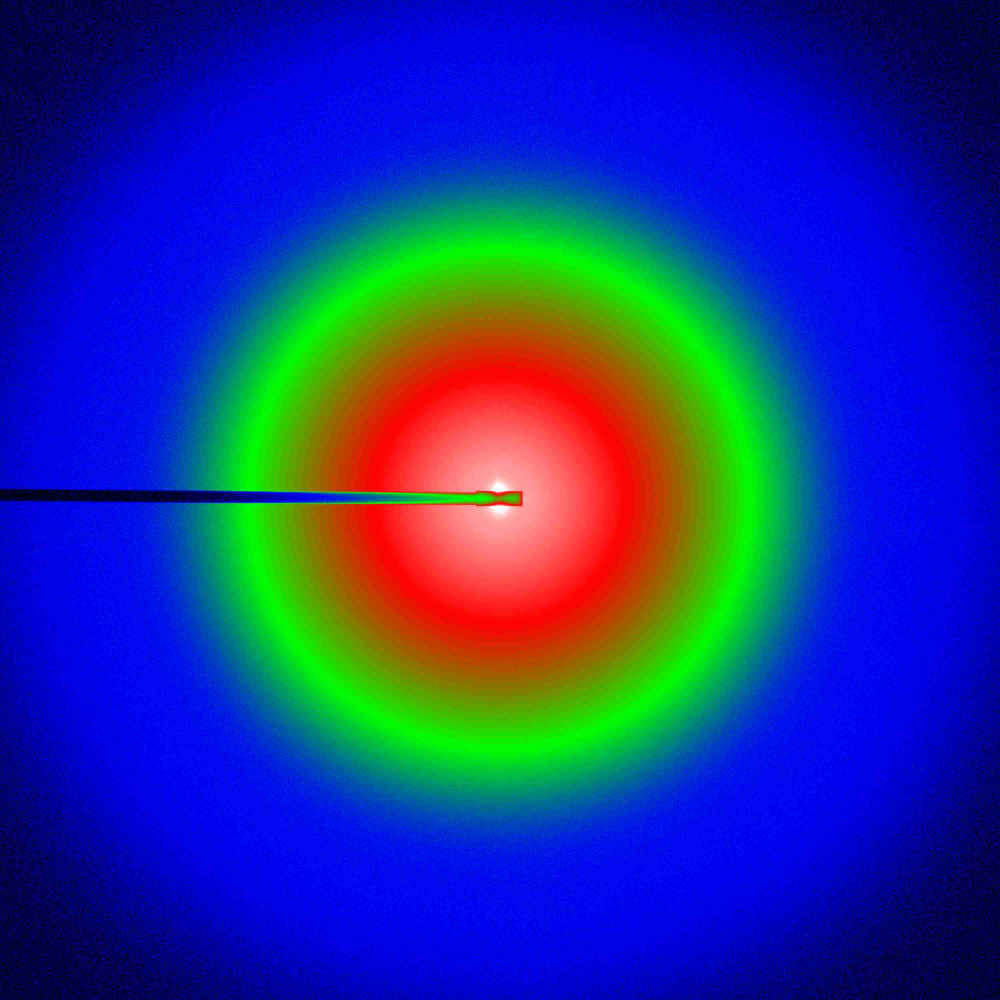}\hfill
  \includegraphics[width=0.2\linewidth]{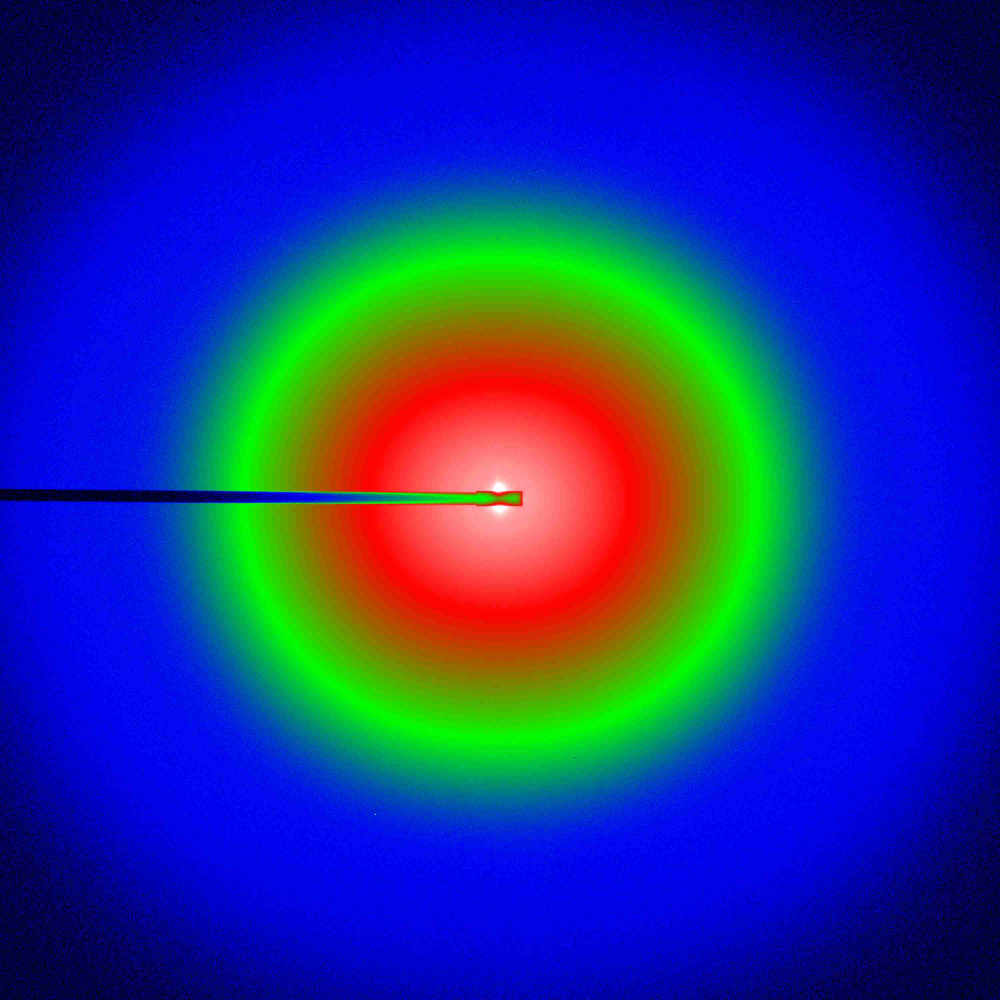}\hfill
  \includegraphics[width=0.2\linewidth]{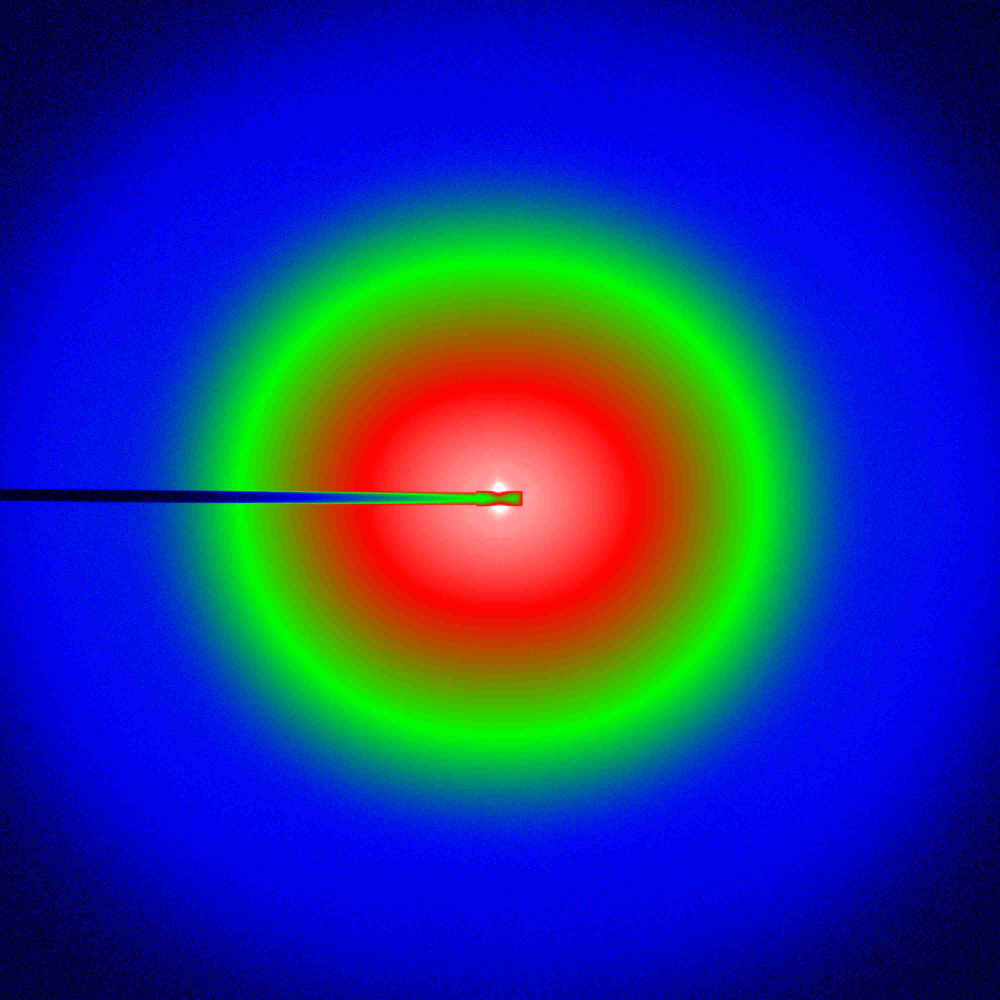}\hfill
  \includegraphics[width=0.2\linewidth]{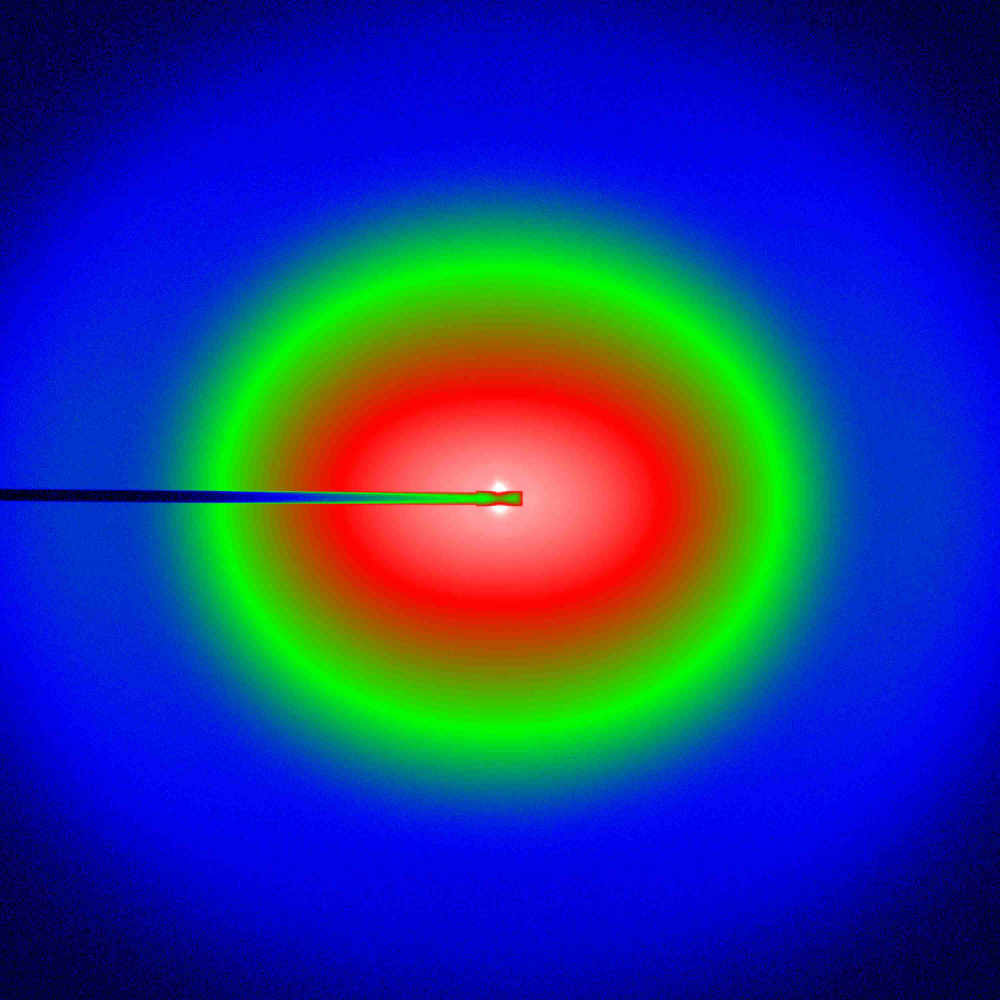}\hfill
  \includegraphics[width=0.2\linewidth]{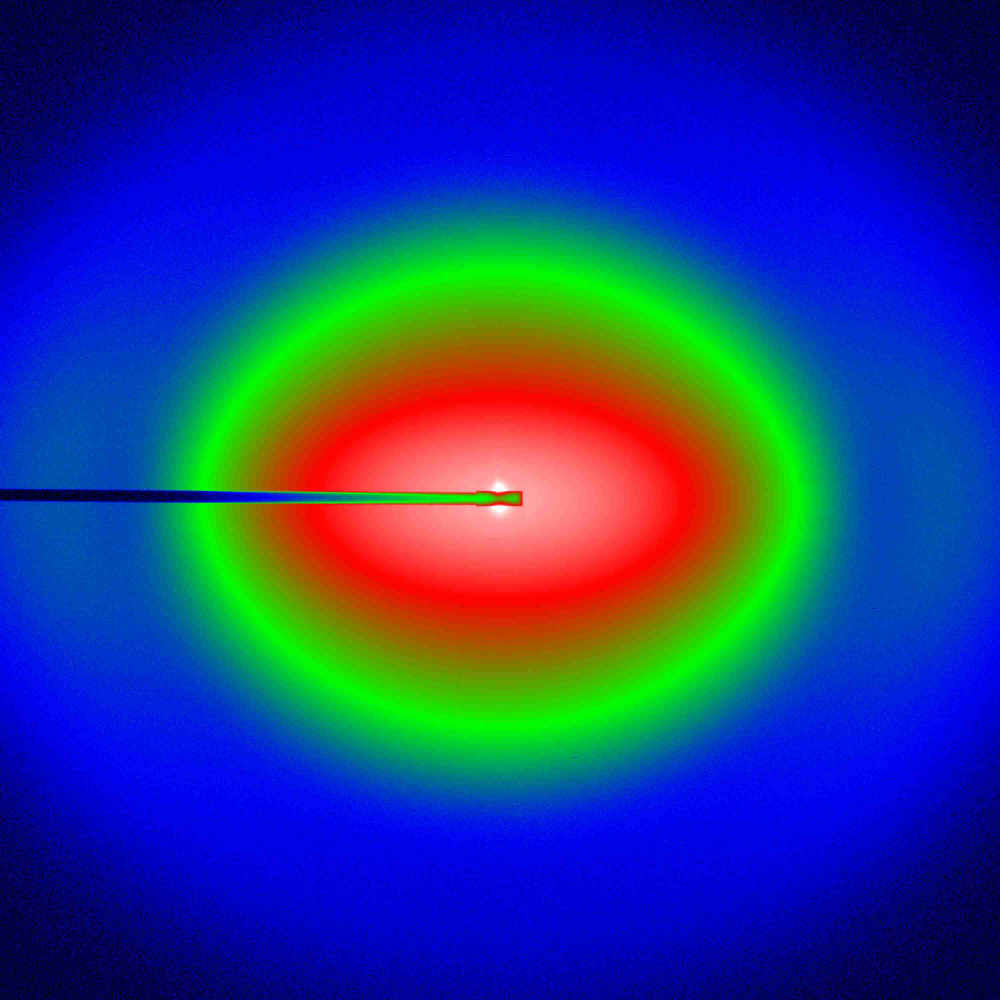}}
  \caption{False color representation of the Small Angle X-ray Scattering resulting from aqueous suspensions  (upper row) and composites (lower row) containing identical number densities of hematite particles. The magnetic flux density with horizontal field direction  increases from left to right ($B_1= 3.2\,\,\rm{mT}$, $B_2= 10.0\,\,\rm{mT}$, $B_3= 98\, \,\rm{mT}$, $B_4= 1.00\,\,\rm{T}$, $B_5= 1.50\,\,\rm{T}$).}
  \label{fig:false_color}
\end{figure}

In order to investigate particle-matrix interactions the ODF of hematite spindles, both, in a hydrogel matrix and, at the same number density, in aqueous suspensions are compared.  In order to achieve neglectable particle-particle interactions, composites with a volume fraction corresponding to an ellipsoid of revolution as small as  $\varrho^*\pi/6=\varrho\sigma_{\rm eq}^3\pi/6=1.57\times 10^{-4}$ are prepared. Here, $\sigma_{\rm eq}$ denotes the equatorial diameter of the spindles. This corresponds to a volume fraction $\varphi_{\rm HEM}\approx 10^{-3}$ of hematite spindles. The particle-matrix interactions are tuned by the polymer volume fraction $\varphi$ and the crosslinking ratio $\chi$ of the hydrogel matrix.
The ODFs are determined from the anisotropy of the 2d SAXS intensity in presence of an external field with varying flux density and orientation with respect to the scattering vector $\mathbf{Q}$.  Due to the low electron density of the hydrogel and water compared to 
hematite, the scattered intensity mainly results from the embedded particles, as visible from the comparison of scattered intensity from composites and pure hydrogel matrix in Fig. A1 in the supplementary material, where the scattering data of both, composite and pure hydrogel matrix are corrected for a background of water filled capillary.  The scattering data displayed in the following
are corrected for the background of a water filled capillary for aqueous suspension and for that of a hydrogel filled capillary 
in the case of composites. Hence, the scattered intensity displayed solely originates from the particles, either in water or hydrogel.\\
In absence of a magnetic field the scattering patterns resulting from both, composites and aqueous suspensions, are completely isotropic (FIG.\,\ref{fig:false_color}) indicating a random orientation of the hematite spindles. With increasing flux density of an external magnetic field, the particles align with their long axes perpendicular to the field direction. With a scattering geometry using a magnetic field perpendicular to the primary beam, scattering vectors $\mathbf{Q}$ both, parallel and perpendicular to the external field $\mathbf{B}$, can simultaneously be observed.

With increasing flux density $B$, in the scattering patterns resulting from both, hydrogels and aqueous suspensions, an increasing anisotropy is visible. However, due to particle-matrix interactions, the increment of composites' anisotropy is less pronounced than that of aqueous suspensions as visible from false color representations of the scattered intensity (FIG.\,\ref{fig:false_color}).
The field-induced changes on the mesostructure of both, composites and aqueous suspensions, are completely reversible: typically 10 s after removing the external field, the scattering patterns are isotropic again.

\begin{figure}
\centering
 \includegraphics[width=0.7\linewidth]{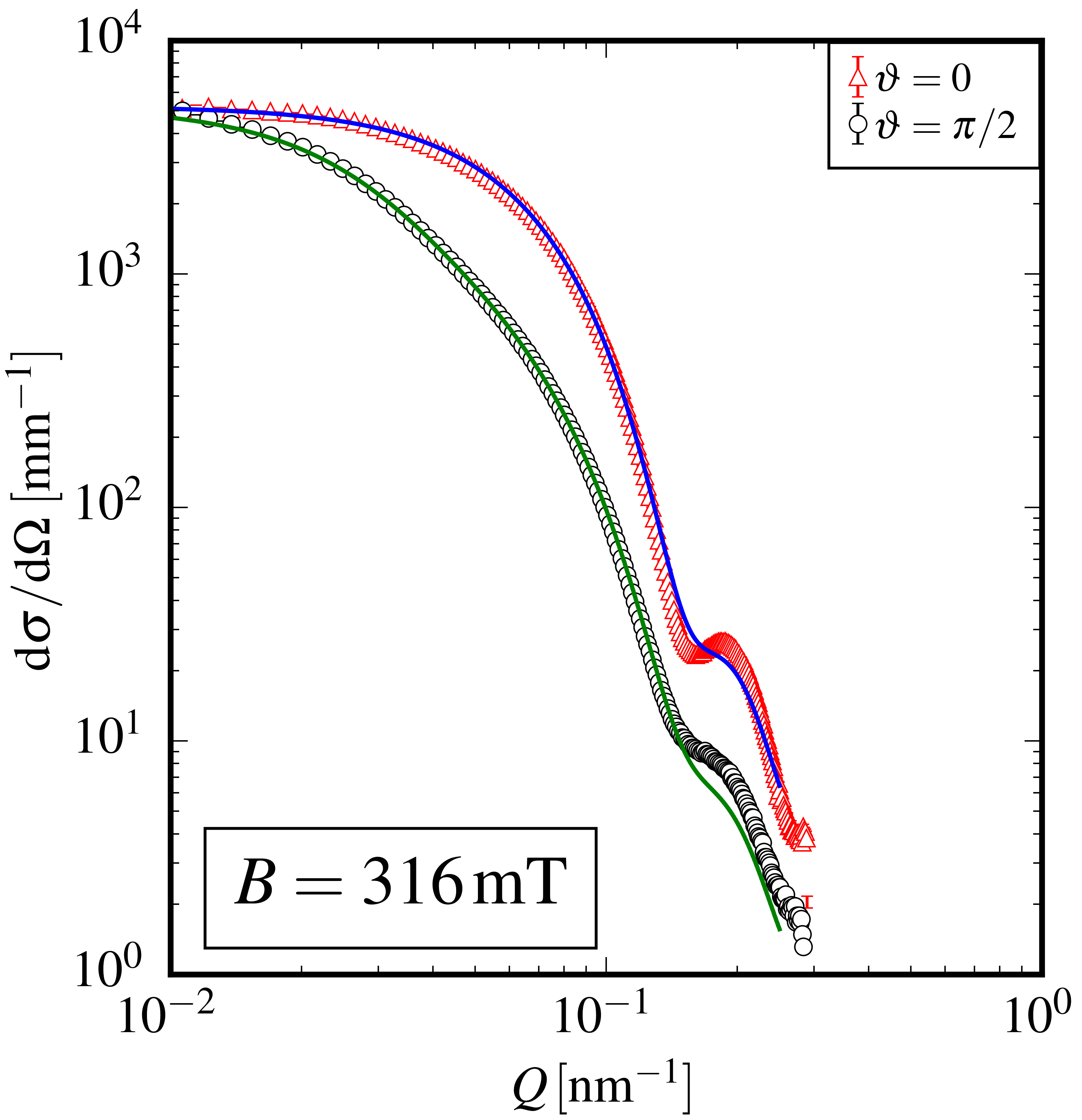}
 \includegraphics[width=0.7\linewidth]{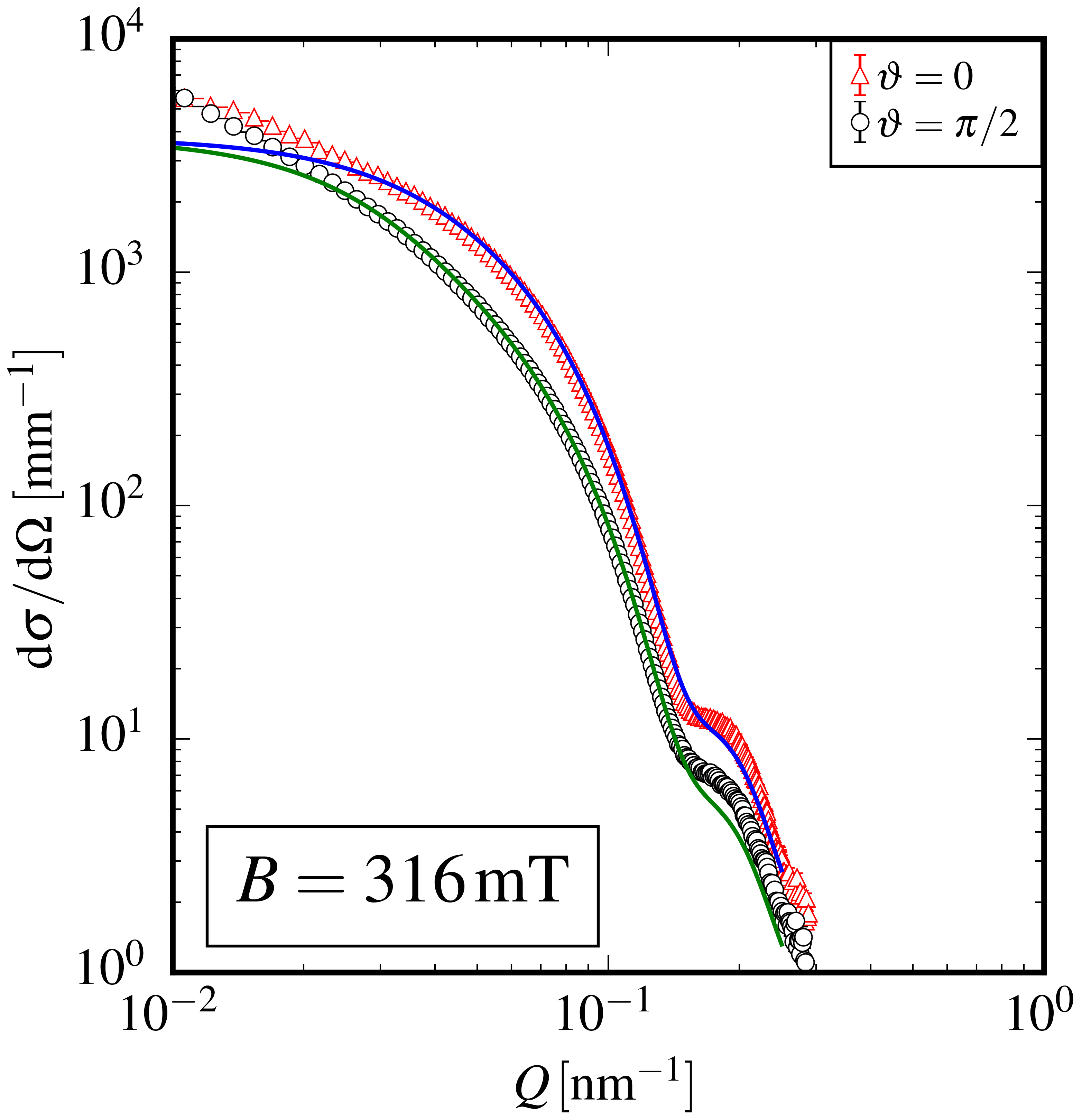}
  \caption{Sector averaged mean intensities of an aqueous suspension of hematite (top) and a hematite-pNIPAM composite (bottom) exemplarily displayed for the scattering vector parallel ($\vartheta_{\rm Q} = 0$)(red triangles)  and perpendicular ($\vartheta_{\rm Q} = \pi/2$)(black circles) to the external magnetic field. The solid lines are the results of a simultaneous least-squares fit of the complete reduced data set with $0 \le \vartheta_{\rm Q}\le \pi/2$ employing a model of Schulz-Flory distributed spindles in an external field. }
  \label{fig:intensity}
\end{figure}

The raw intensities detected with a 2d CCD detector (3840 x 3840 pixels)  are pixelwise corrected for parasitic scattering, electronic background and detection efficiency.   Subsequently, the detector plane is partitioned  into 72 sectors with an acceptance of $\Delta\varphi = \pm 2.5^\circ$ defining the direction of the scattering vector $\mathbf{Q}$ with respect to the external field. 
The modulus $Q$ is related to a pixels' distance to the center of the scattering pattern. Therefore the detector is  additionally partitioned  in annuli of the width  $\Delta r=2.48\times 10^{-4}\,{\rm m}$ corresponding to the diagonal of four pixels.
Due to the cylindrical symmetry of the particles for the form factor the relation
\begin{align}
\label{eq:symmetry_relation}
P(Q,\vartheta_{\rm{Q}})&= P(Q,-\vartheta_{\rm{Q}})\notag\\
&= P(Q,\pi-\vartheta_{\rm{Q}}) \notag\\
&= P(Q,\pi+\vartheta_{\rm{Q}})
\end{align}
results, 
where $\vartheta_{\rm{Q}}$ denotes the angle between the scattering vector $\bf{Q}$ and the particle director $\hat{\mathbf{u}}$.  As a consequence, only 19 of 72 sectors, which are averages of four (respectively two for $\vartheta_{\rm Q}=0$ and $\vartheta_{\rm Q}=\pi/2$),  are symmetrically independent. The data reduced employing the symmetry relation \eqref{eq:symmetry_relation} are used for the numerical analysis of scattering data.\\
In absence of an external magnetic field the orientation of the hematite spindles is random irrespective of the surrounding medium, either water or hydrogel. The scattering pattern resulting from such a spherically symmetric ensemble is radially symmetric, too. In this case, all sector averages are equal within experimental accuracy, independent of the angle $\vartheta_{\rm Q}$. In presence of a magnetic field, due to the interaction with the particles' permanent and induced magnetic dipoles, the particles align. As a consequence, the spherical symmetry is reduced to a cylinder symmetry as visible, both,  in the 2d false colour representation of the scattered intensity (FIG.\,\ref{fig:false_color}), and the angular dependence of the sector averages (FIG.\,\ref{fig:intensity}). With increasing anisotropy of the ODF the differences between sectors, maximum between directions  parallel and perpendicular to the external field, increase as displayed in FIG.\,\ref{fig:intensity} for a flux density of $B= 316\,{\rm mT}$ resulting from particles in aqueous suspension and in a hydrogel matrix.\\
It is obvious that differences between $\vartheta_{\rm{Q}}=0$ and $\vartheta_{\rm{Q}}= \pi/2$ are more pronounced for the aqueous suspensions than for the pNIPAM-composites. While at $316\;\rm{mT}$ the particles in the hydrogel are still more or less statistically aligned, in water, the particles are nearly completely aligned perpendicular to the external field. \\
Both, for composites and aqueous suspensions, the scattered intensity in the direction $\mathbf{Q}\parallel \mathbf{B}$ ($\vartheta_{\rm Q}=0$) is more structured than for $\mathbf{Q}\perp\mathbf{B}$ ($\vartheta_{\rm Q}=\pi/2$).
Though the hematite spindles align with their long axis perpendicular to the external magnetic field 
a rotation around their short axis is still possible. Hence, the electron density for a particle rotating around its short axis resembles a disc with its' equatorial plane always perpendicular to the field direction and, as a consequence, $\mathbf{Q}$ always perpendicular to the particle director.  Perpendicular to the field, however, a circular distribution of particle directors over $2\pi$ is observed, leading to an average orientation of the particle director with respect to $\mathbf{Q}$, smearing out the scattered intensity for $\vartheta_{\rm Q}=\pi/2$. Applying a magnetic field parallel to the primary beam, as expected, a symmetric scattering pattern is observed.

The SAXS data are analysed assuming homogeneous, spindle-shaped particles with a Schulz-Flory size distribution of their equatorial radii. For simplicity of the model, a constant aspect ratio $\nu$ is assumed. Since the volume fraction of hematite spindles is typically $\varphi_{\rm HEM}=10^{-3}$ and the interparticle distances are significantly larger than the particle size, particle-particle interactions can be neglected. As a consequence, with a structure factor $S(Q)\equiv 1$, the scattering data can solely be described by the form factor $P(Q,\vartheta_{\rm Q})$ of an ensemble of spindles underlying both, a size distribution, and an orientational distribution (FIG.\,\ref{fig:intensity}). With respect to the models' simplicity, the experimental data are described very well as a function of both, modulus and direction of $\mathbf{Q}$.  Small deviations at $|\mathbf{Q}|\approx 0.2\,{\rm nm}^{-1}$ corresponding to length scales of $d\approx  30\,{\rm nm}$ in real space can be attributed to deviations of the real particle shape from the spindle model and neglecting a distribution of aspect ratios. Deviations at very small scattering vectors in the hydrogel composites result from weak aggregation of the hematite particles. Since the long axis of hematite spindles used in these experiments is typically $l=\nu r_{\rm eq}\approx 200\,{\rm nm}$, the orientation of the particles mainly influences the scattering function at $|\mathbf{Q}|\approx 2\pi/(200\, {\rm nm})$. In the latter $|\mathbf{Q}|$-range the model accurately describes the scattering data.

\begin{figure}
\centering
  \includegraphics[width=0.8\linewidth]{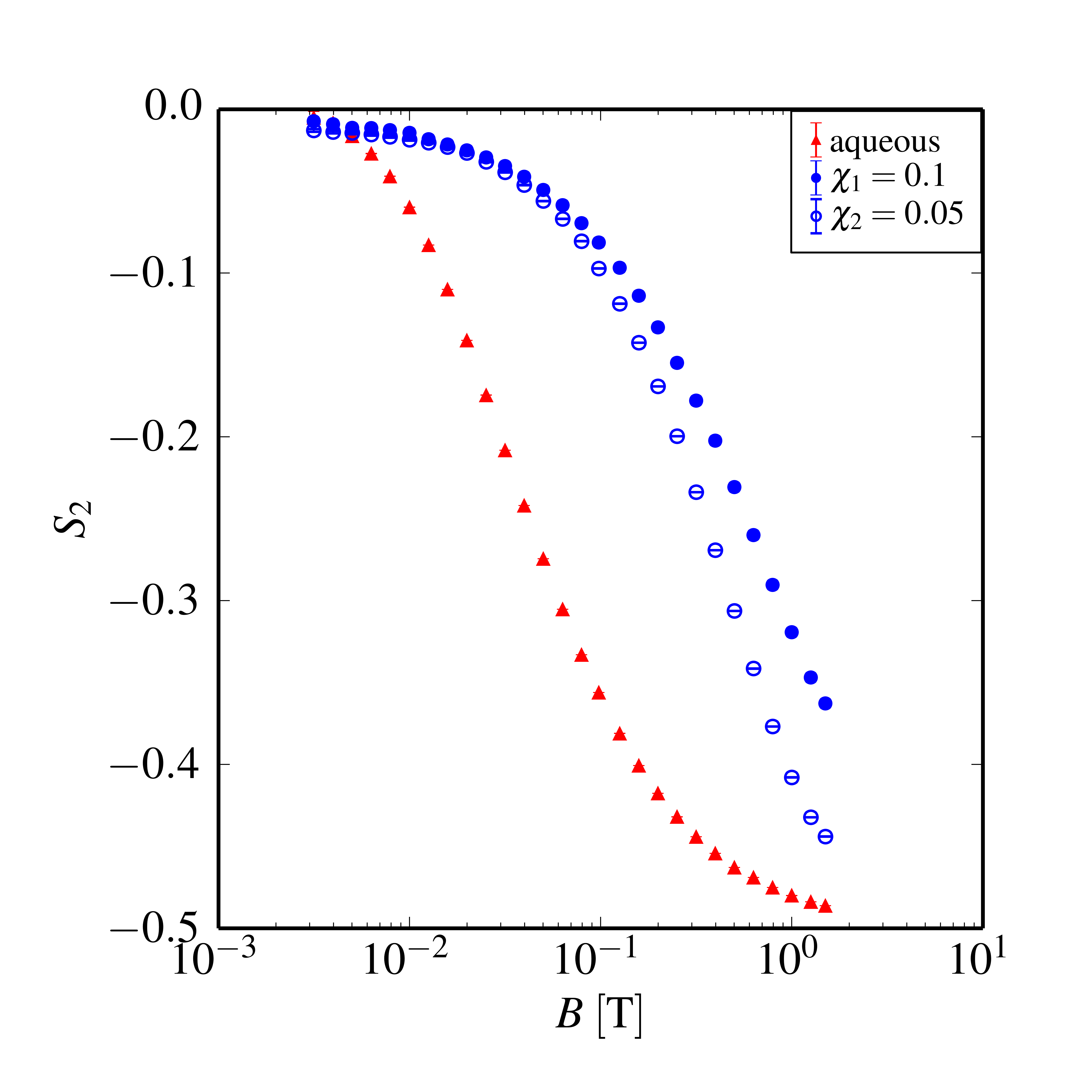}
  \caption{Order parameter $S_2$ of an aqueous hematite suspension ($\nu=4.0$) (red triangles) and hydrogel composites with constant volume fraction $\varrho^*\pi /6=1.57\times 10^{-4}$ of hematite. Both hydrogels contain an identical polymer  volume fraction ($\varphi=0.05$) at different crosslinking ratios  $\chi=0.05$ (open blue circles) and $\chi=0.1$ (filled blue circles). }
  \label{fig:S2crosslink}
\end{figure}

\begin{figure}
\centering
  \includegraphics[width=0.8\linewidth]{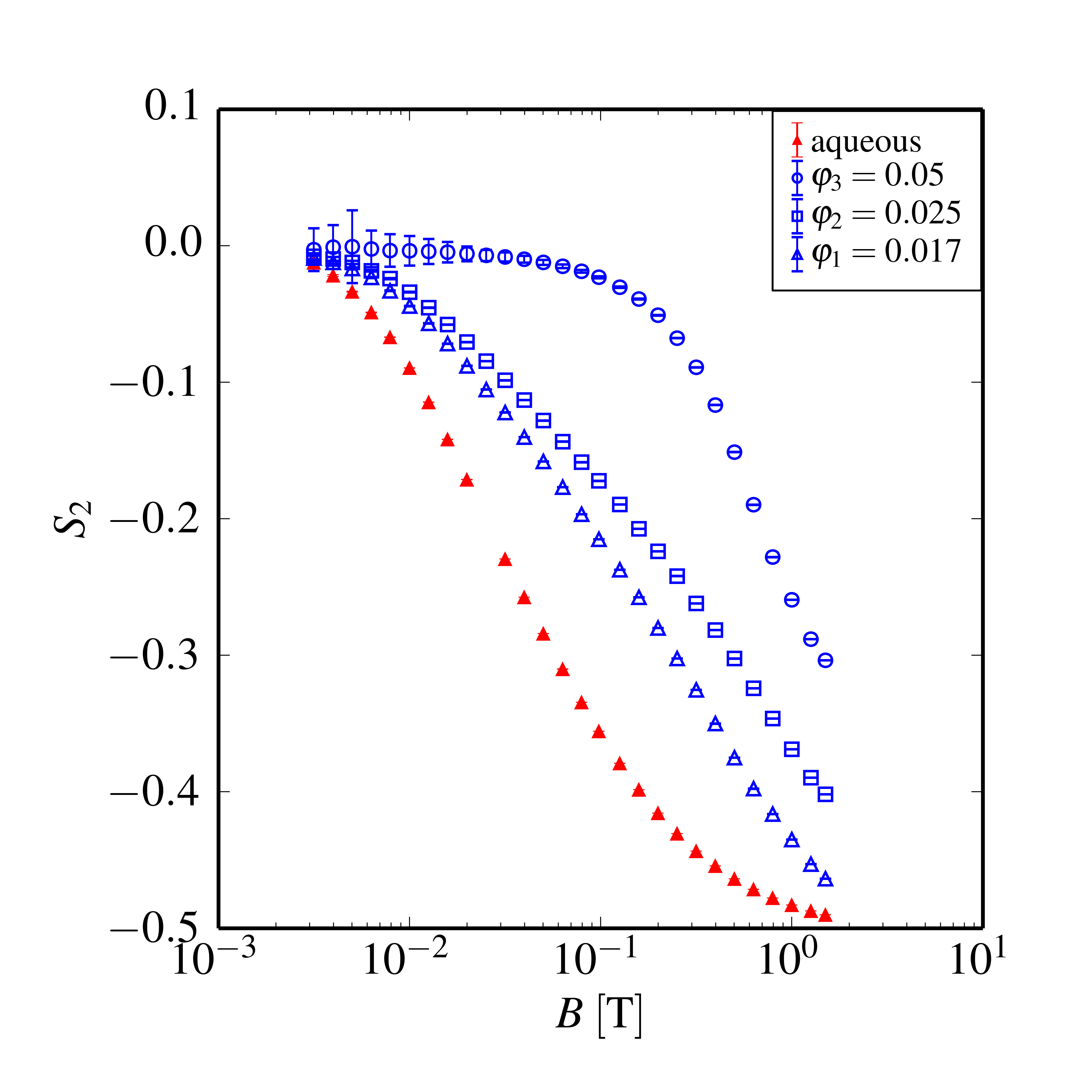}
  \caption{Order parameter $S_2$ of an aqueous hematite suspension (filled red triangles) with an aspect ratio of $\nu=4.0$ and hydrogel composites of different volume fractions ( $\varphi_1=0.017$, $\varphi_2=0.025$, $\varphi_3=0.050$) (open blue symbols) at identical crosslinking density $\chi=0.05$ and identical
  volume fraction of $\varrho^*\pi/6=1.5\times 10^{-4} $.}
  \label{fig:S2volume}
\end{figure}

Employing the Boltzmann approach \eqref{eq:Boltzmann_ODF} for the ODF, the nematic order parameter $S_2$ \eqref{eq:s2_nematic} is obtained by canonically averaging the distribution function. $S_2$ quantifies the field-induced alignment of the particles in dependence on the flux density of the applied field.  In FIG.\,\ref{fig:S2crosslink} and FIG.\,\ref{fig:S2volume} the field-dependent order parameters of aqueous suspensions and hydrogel-composites containing identical number densities of hematite spindles are compared. Both, for aqueous suspensions and hydrogel-composites, a field-induced isotropic-nematic transition is observed, although in the hydrogel-composites the nematic alignment is less pronounced than in aqueous suspensions.\\
In FIG.\,\ref{fig:S2crosslink} the influence of the crosslinking ratio $\chi$ on the particle-matrix interactions is visible. In hydrogels with identical
polymer volume fraction $\varphi$, the nematic alignment is progressively hindered with increasing crosslinking ratio $\chi$. Increasing crosslinking ratios result in increasing elasticity of the hydrogel matrix causing restoring torques to the magnetically induced angular displacement.
 
In FIG.\,\ref{fig:S2volume}, the influence of the polymer volume fraction $\varphi$ at constant crosslinking ratio $\chi$ is displayed.
An increasing polymer volume also induces a rising elasticity of the hydrogel leading as well to a progessively hindered nematic alignment 
of the magnetic spindles. Both, in aqueous suspensions and hydrogel composites, an equilibrium mesostructure is obtained within several seconds and the alignment is completely reversible.  This clearly indicates that the hindered alignment is a result of the viscoelasticity of the composites: purely viscous friction would  prolong the equilibration but result after infinite time in the same time-averaged mesostructure.

\subsection{Rheology}

The macroscopic viscoelastic properties of these ferrogels are investigated by means of oscillatory shear experiments in presence of an external magnetic field. In oscillatory, opposite to rotational shear experiments, defined, small deformations $\gamma$ of the samples can be applied, which avoid shear induced structural changes of the hydrogel matrix. In addition, the frequency dependent, complex shear moduli $G'$ and $G''$, characterizing the elastically stored and viscously dissipated energy density, are obtained. The intersection of the storage modulus $G'$ and the loss modulus $G''$ at the phase angle $\tan\delta=G''/G'=1$ indicates a transition from solid state to liquid state properties of viscoelastic systems. Influences of field-induced mesostructural changes on the viscoelastic properties of ferrogels are investigated by variation of the flux density of a magnetic field parallel to the shear gradient. 

\begin{figure}
\centering
 \includegraphics[width=0.82\linewidth]{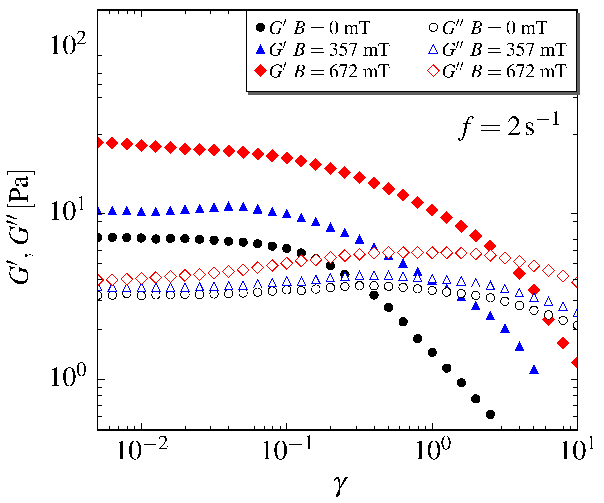}\hfill
  \caption{Storage- and loss moduli $G'$ (filled symbols) and $G'' $(open symbols) of a hemattie-pNIPAM composite in dependence on the deformation $\gamma$. At a constant frequency of $f= 2\,\rm{s}^{-1}$ the flux density of an external field changes from $B=0\,\,{\rm mT}$ to $B=677\,{\rm mT}$.}
  \label{fig:rheo_1}
\end{figure}

In an external magnetic field with increasing flux density, the rotational mobility of the shape anisotropic particles is progressively confined, leading to a drastic increase of the elastic modulus $G'$, especially at small deformations. The loss modulus $G''$, however, is nearly independent of the flux density (FIG.\,\ref{fig:rheo_1}). Hence, increasing flux densities lead to a significant increase of the elastically stored energy, whereas the energy dissipated by viscous friction is, especially at small deformations, independent of an external magnetic field.  Phenomenologically,  the hematite spindles act as additional, field-dependent crosslinkers within the polymer matrix.

\begin{figure}
\centering
  \includegraphics[width=0.8\linewidth]{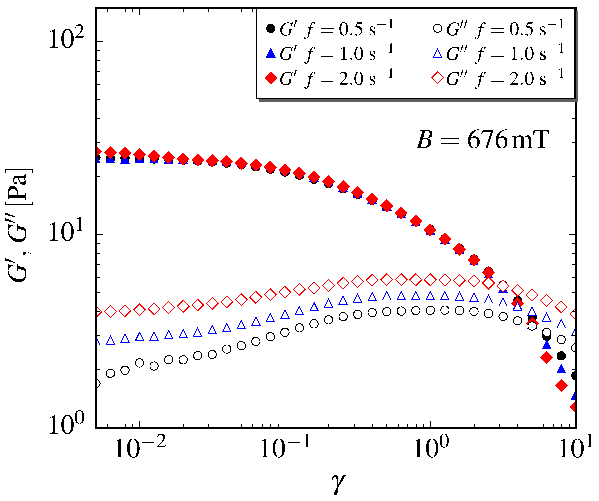}
  \caption{Storage- and loss moduli $G'$ (filled symbols) and $G'' $(open symbols) of a hematite-pNIPAM composite in dependence on the deformation $\gamma$. At a constant flux density of $B=676\,{\rm mT}$ the oscillatory frequency is changed from $f=0.5\,{\rm s}^{-1}$ to $f=2.0\,{\rm s}^{-1}$.}
  \label{fig:rheo_2}
\end{figure}

For timescales $t \gg f^{-1}$, the time-averaged stored energy density is in first approximation independent of the frequency. An increased frequency, however, results in an increase of the dissipated energy density per time, as visible in FIG.\,\ref{fig:rheo_2}, where the loss modulus $G''$ increases with the oscillation frequency, while the storage modulus $G'$ does only weakly depend on the frequency.

\section{Conclusions and Outlook}
Due to both, permanent and, as a consequence of their negative magnetic anisotropy induced dipole moments,  a torque is exerted to anisotropic hematite particles in presence of an external magnetic field. The field-induced rotation of particles embedded in a hydrogel leads to an elastic deformation of the polymer network causing a restoring torque related to the elastic modulus of the hydrogel. Hence, the orientational distribution function of hematite spindles as a function of the flux density is determined by particle-matrix interactions in these composites. As a reference system, identical particles suspended at identical number density in a Newtonian fluid such as water, allows in absence of restoring torques to determine the ODF in thermal equilibrium.  The interactions of the particles with an external magnetic field provoke a field-induced isotropic-nematic phase transition which is progressively hindered with increasing elasticity of the matrix. 
The orientational correlation of the particles is quantified by the nematic order parameter $S_2$ in dependence on the flux density.  Since the elastic moduli of hydrogels are influenced, by the polymer volume fraction $\varphi$ as well as the crosslinking ratio $\chi$, both parameters affect the nematic order parameter $S_2$ of hematite spindles in different hydrogel matrices as a function of the flux density as visible in FIG.\,\ref{fig:S2crosslink} and FIG.\,\ref{fig:S2volume}.

The complete reversibility of the field-induced nematic transition clearly indicates that the hindrance of nematic alignment is related to the elasticity and not to an increased Newtonian viscosity: purely viscous 
particle-matrix interactions would not change the thermal  equilibrium at a given flux density, but only slow down the systems' response to a jump in magnetic flux density. For all samples, the thermal equilibrium is reached after several seconds and removing of the external field results after few seconds in isotropic ODFs.

Due to the elastic particle-matrix interactions as a result of the field-induced rotation of anisotropic particles, these particles act as microrheological probes. In shear experiments, in an opposite way, the rotation of particles is induced by the vorticity of shear flow. Freely rotating particles minimize their rotational friction by following the angular velocity field of their environment.
In presence of a magnetic field, however, the rotational mobility of the particles  is progressively confined with increasing flux density. Also in oscillatory shear experiments, at least for small deformations, an elastic particle-matrix interaction is observed: increasing flux density solely affects the elastic modulus $G'$, whereas the loss modulus for deformations $\gamma< 10^{-1}$ is nearly unchanged at constant oscillation 
frequency $f$. 
With the flux density increasing rotational arrest of hematite particles leads as increasing polymer volume fraction $\varphi$ and crosslinking ratio $\chi$ to rising elasticity of the composites. Phenomenologically, embedded hematite particles act as additional, field-dependent cross linkers.

As expected, the time-averaged elastically stored energy and therewith the elastic modulus  is independent of the frequency $f$, if the average is taken for times $t\gg f^{-1}$. On the other hand, the energy dissipated by viscous friction is proportional to the angular velocity $\omega=2\pi f$. As expected, at constant flux density, the loss modulus increases with the oscillation frequency 
$f$ as displayed in FIG.\,\ref{fig:rheo_2}. The increase of loss moduli is nearly independent of the deformation $\gamma$. 

The macroscopic rheological properties of ferrogels containing anisotropic, magnetic particles can be influenced by the stimulus of an external magnetic field. This macroscopic change of rheological properties is
caused by a field-induced change of the composites' mesostructure as quantified by field-dependent nematic order parameters. Further insights into the shear-induced changes of the mesostructure are expected from
simultaneous rheological and SAXS-experiments.

With known magnetic properties of the hematite spindles, i.e. the magnitude of their magnetic moment, the direction of the magnetic moment with respect to the particle axis and their magnetic anisotropy, the torque 
acting on the hematite spindles can be estimated. On the other hand, from the particles geometry and their roatation, the deformation of the hydrogel network is accessible. The quantitative analysis of the hydrogels' elastic properties from the field-dependent nematic order parameter is the scope of future work. First attempts to model the mesostructural response of
ferrogels are described  and Weeber et al.\cite{Weeber2015} and Pessot et al.\cite{Pessot2016} by molecular dynamics on the coarse grained level. The hindrance
of nematic alignment of anisotropic particles due to elastic distortion of gels can be used to compare these theoretic results 
with experimental data, since geometry and torques acting on the particles as well as the elastic properties of the matrix are known. 

Furthermore, Brader et al. \citep{Brader2010} proposed a mode coupling approach to describe the nonlinear viscoelastic behaviour of colloidal particles. As an input for this quantitative approach the static structure factor of the colloidal system is used. Since our hydrogel consists of chemically crosslinked polymer
spheres, this approach is also promising to analyse the nonlinear viscoelasticity of these composites. Due to the limited $Q-$range accessible by light scattering and the small electron density of the polymer, additional
neutron scattering experiments are required to access the structure factor of the hydrogel matrix in the relevant $Q-$range.

\section{Acknowledgements}

We acknowledge the European Synchrotron Radiation Facility (ESRF) for providing beamtime at the beamlines ID02 and ID10 and the Deutsche Forschungsgemeinschaft for financial support within the priority program SPP 1681. We thank Beatrice Ruta, Sylvain Prevost and Johannes M\"{o}ller for support during the experiments at ESRF and the Center for Electron Microscopy (EMZ) of the Universit\"{a}tsmedizin Rostock for the opportunity to take the TEM micrographs. \\

\bibliographystyle{aipauth4-1}
\bibliography{JCP}% Produces the bibliography via BibTeX.

\end{document}